\documentclass[11pt,a4paper]{article}
\usepackage[T1]{fontenc}
\usepackage[utf8]{inputenc}
\usepackage{lmodern}
\usepackage[margin=25mm]{geometry}
\usepackage{amsmath,amssymb,graphicx,booktabs,array,longtable}
\usepackage{caption,placeins,microtype}
\usepackage{xcolor}
\usepackage[colorlinks=true,linkcolor=blue!45!black,citecolor=blue!45!black,urlcolor=blue!45!black]{hyperref}
\usepackage{xurl}
\hypersetup{pdfauthor={Ali Hosseinzadeh; A. Hosseiny; Reza Jafari}}
\newcommand{\StudyReleaseURL}{https://github.com/alihosseinzadehph/polymarket-participation-replication/tree/81f5a4940be3a2d56bccb90bb3f3042af0f10086}
\title{Political Orientation and Participation in Prediction Markets: Large Stakes Converge, Small Stakes Diverge}
\author{Ali Hosseinzadeh$^{1}$, Ali Hosseiny$^{1}$, Reza Jafari$^{1}$\\[5pt]
\small $^{1}$Department of Physics, Shahid Beheshti University,\\
\small Evin, Tehran 1983969411, Iran\\
\small Ali Hosseinzadeh,
\href{mailto:ali_hosseinzadeh@sbu.ac.ir}{\nolinkurl{ali_hosseinzadeh@sbu.ac.ir}}\\
\small Ali. Hosseiny:
\href{mailto:al_hosseiny@sbu.ac.ir}{\nolinkurl{al_hosseiny@sbu.ac.ir}}\\
\small Reza Jafari:
\href{mailto:g_jafari@sbu.ac.ir}{\nolinkurl{g_jafari@sbu.ac.ir}}}
\date{}

\date{}

\begin{document}
\maketitle

\begin{abstract}

Prediction markets offer a way to study collective expectations and beliefs: people express their views on various events and, by putting money behind their predictions, take on the risks of their judgments. As a result, the scale of investment can reveal additional dimensions of individual behavior and beliefs. In this study, we use a complex network approach to examine the structure of user participation, drawing on roughly 1.14 billion transaction records from the Polymarket platform, covering more than 3.1 million wallet addresses and 764,000 prediction events. To do this, we build a bipartite network of users and events and analyze patterns of investment concentration, connections between topics, and the network's community structure. Our findings show a highly skewed distribution of financial participation: 1\% of events account for 64\% of total transaction volume, while 1\% of wallets account for 79.2\% of total financial participation. We also examine how investment scale and political orientation shape user participation patterns. Focusing on the 2024 U.S. presidential election, we compare users who invested exclusively in a Trump victory or exclusively in a Harris victory, grouping wallets by purchase size rather than by retail or institutional status. The results show that among large-scale investors, capital allocation patterns across topics are more similar between the two political groups; in contrast, these patterns diverge more and vary more among small-scale investors. This result holds across six different ways of representing and comparing the networks, and four of these differences are also statistically supported.
\end{abstract}

\section{Introduction}

Prediction markets turn people's positions on future events into market prices. This gives us a way to aggregate information and see how collective expectations form. Unlike polls or free opinions, users here put money behind their predictions and accept the financial risk. Putting money at risk distinguishes a trade from a stated opinion. Prediction markets can aggregate information \cite{wolfers}, but participants need not trade for the same reason or follow the same strategy \cite{rothschildsethi}.

Polymarket, as one of the largest decentralized prediction platforms, lets pseudonymous addresses trade contracts on everything from politics and macroeconomics to sports and cryptocurrency. On these platforms, executing a trade is distinct from minting, converting or settling contracts \cite{tsang,sokdpm}; our volume measure counts executed trades. Recent studies have looked at arbitrage on Polymarket, election-market accounting, market microstructure, political shocks, profits, and measures of political leaning \cite{saguillo,tsang,tsangshocks,qin,akey,chenweb3}. Other work asks whether prediction-market accuracy reflects crowd wisdom or a smaller set of skilled accounts \cite{gomezcram,luongheesen}. These findings motivate tests of concentration and network organization, but they do not by themselves justify inferring skill, private information, or political identity from network centrality.

In this study, we examine the structure of user participation across the whole platform using a complex-network approach. Drawing on roughly 1.14 billion transaction records, more than 3.1 million wallet addresses, and 764,000 prediction events, we construct a bipartite network in which each wallet is linked to the events it traded, with each link weighted by that wallet's total trading amount in the event. This representation lets us analyze patterns of investment concentration, connections between topics, and the community structure of the network. The core questions are: where does activity concentrate across subjects and contracts? Which topics share wallets? And is the subsequent subject allocation more similar between candidate-linked high-purchase cohorts than between the corresponding low-purchase cohorts?

Our initial expectation is that in political events, political orientation shapes how people understand events and, consequently, their investment patterns. However, in financial markets there is no single "winning" strategy: one group or firm researches and decides on a sell strategy for a stock, while another group or firm pursues a buy strategy for the same stock, and it is precisely this difference in strategies that makes trading possible. In models of market ecology, the return to a trading strategy depends on how capital is distributed across competing strategies \cite{farmer2002ecology,scholl2021ecology}. Yet at the large-investment scale, we observe a kind of trend formation, and one could argue that compared with small-scale investment, there is less diversity in investment strategy. On this basis, the main hypothesis of the study is that large investors converge in their capital allocation beyond the election market and form a kind of consensus and trend, while small investors have more diverse strategies and act more under the influence of their own political bias. This hypothesis is consistent with the idea of "convergence of costly beliefs at large scale": the larger the amount involved, the higher the cost of bias, and therefore allocation moves toward more efficient and more similar patterns.

To test this hypothesis, we quantify event- and wallet-level concentration and assess whether the tails of monetary and degree distributions are consistent with a power law, following the procedure of Clauset, Shalizi and Newman \cite{clauset}. A power-law fit to the upper tail of the number of wallets per event gave an exponent of about 2.4 (Supplementary Table~S5). We then analyze the complete wallet--event graph under monetary and wallet-normalized weights and compare shared subject participation with a degree-constrained bipartite null. Finally, September 2024 purchases in the presidential-election winner market define candidate-linked purchase-size strata, whose non-focal subject allocation is measured during October. This case study asks whether the cross-subject allocation of high- and low-purchase wallets differs across candidate outcomes. The contribution of the study lies in combining three layers: an analysis of concentration and power-law distributions across the whole platform, community detection in the complete wallet--event network, and an empirical test of the convergence-of-large-investors versus divergence-and-bias-of-small-investors hypothesis in the setting of a real election event.

\section{From trades to the wallet--event network}
\label{sec:participation-data}

\subsection{Trade records and event mapping}
On Polymarket, a single event can contain several markets. The markets for a Harris victory and a Trump victory, for example, belonged to the same 2024 US presidential election winner event. A wallet that traded in both has one link to that event; trades in either market contribute to its weight.

Executed trades (fills) came from TimeSeventeen's cleaned, blockchain-derived Polymarket-v1 and Polymarket-v2 datasets on Hugging Face \cite{qin,source,sourcev2}. The v1 records span 21 November 2022--28 April 2026 and v2 spans 3 April--15 August 2026; we retained both exchange generations during their calendar overlap. The snapshots frozen on 1 September 2026 contain 1,143,528,694 fills totaling approximately 41.309 billion USDC. Supplementary Section~S1 and Table~S1 identify the trade files and their public versions.

We matched each market's condition identifier to its parent event using a 1 September 2026 snapshot of open and closed events from Polymarket's Gamma API \cite{gammaapi}.\footnote{We paginated \texttt{GET /events/keyset} from 11:40:20 to 12:54:17 UTC. Of 2,948,008 market records, 1,380 lacked a condition identifier.} Titles and tags reflect that snapshot, not necessarily the metadata visible at trade time. We calculated event volumes from matched fills rather than using the platform's lifetime volume figures.

For the leaderboard analyses only, we retrieved two all-time top-1,000 lists from Polymarket's Data API on 4 September 2026, one ranked by reported volume and the other by reported profit \cite{leaderboardapi}. These ranks could change during pagination and do not define the wallet population used to build the network.

\subsection{Wallet participation and event subjects}
Each mapped fill counts once toward event volume. For wallet participation we count its amount for both the maker and the taker: total participation in mapped events is therefore twice their fill volume. This measures gross trading by address, not net investment or holdings. Wallet concentration uses all 3,106,517 observed addresses and includes activity from the 52,022 fills that could not be mapped to an event. Those fills do not enter event volumes or the network.

The bipartite network links 3,106,514 wallets to 764,378 positive-volume events through 255,704,359 distinct wallet--event pairs, with no rank or activity cutoff. A wallet's degree counts the events it traded; an event's degree counts the wallets that traded it. We weight each link either by the wallet's total traded amount in that event, $w_{ue}$, or by the fraction of its mapped trading allocated there, $w_{ue}/\sum_{e'}w_{ue'}$. The links are the same in both cases, but under the second weighting each wallet contributes total weight one.

We assigned each event one of 21 subjects using its title, the short name in its web address and its platform tags. Rules for contract form and outcomes precede ordered tag and title matches; unmatched events enter Other. Intraday cryptocurrency, for example, requires both a cryptocurrency reference and a short-interval cue. Supplementary Section~S2 and Tables~S3--S4 give the rule order, subject counts and mapping to seven broad families. The classifier is supplied with the code; independent double coding of a stratified sample remains pending.

Subject labels do not enter community detection. We use them separately to construct the subject networks: summing a wallet's trading over events in the same subject gives its allocation to that subject, while recording only whether it traded there lets us count how many wallets two subjects share.

\section{Results}
\subsection{Markets and Participation Are Strongly Concentrated}

\begin{figure}[ht]
\centering
\includegraphics[width=0.58\textwidth]{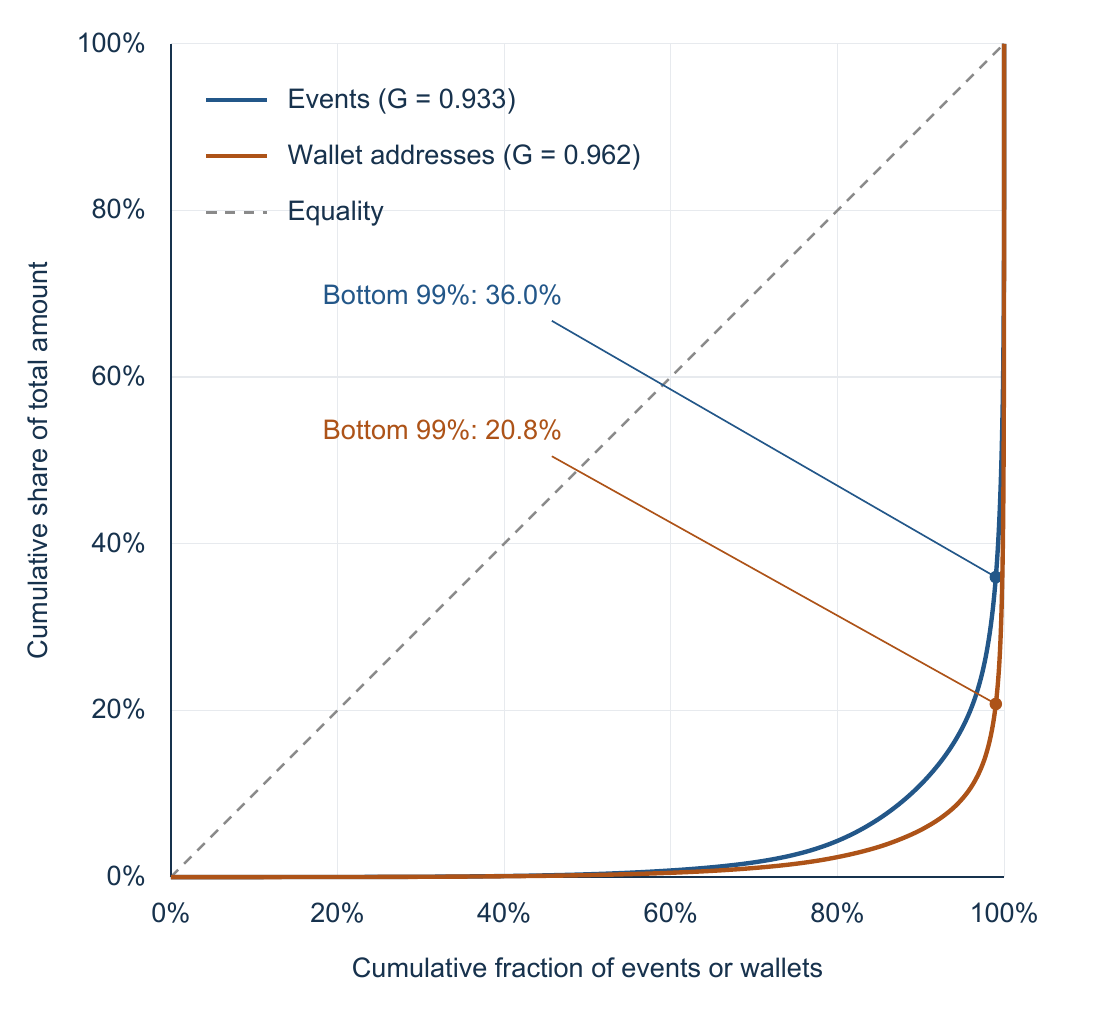}   
\caption{\textbf{Concentration of trading across events and wallets.} Each Lorenz curve orders events or wallets from the smallest to the largest traded amount. The horizontal axis shows the cumulative share of events or wallets; the vertical axis shows their cumulative share of the total amount. The dashed diagonal denotes equal amounts for all events or wallets, and $G$ is the Gini coefficient. Markers indicate the share accounted for by the lowest-ranked 99\%. Event volume counts each trade once; wallet participation counts the amount for both the maker and the taker.}
\label{fig:concentration}
\end{figure}

The largest 1\% of events by volume accounted for 64.0\% of mapped trading volume. Trading was more concentrated across wallets: the largest 1\% by traded amount accounted for 79.2\% of total wallet participation. The Gini coefficients were 0.933 for event volume and 0.962 for wallet participation (Fig.~\ref{fig:concentration}).

\subsection{Pareto scaling is selective rather than universal}
\begin{figure}[ht]
\centering
\includegraphics[width=\textwidth]{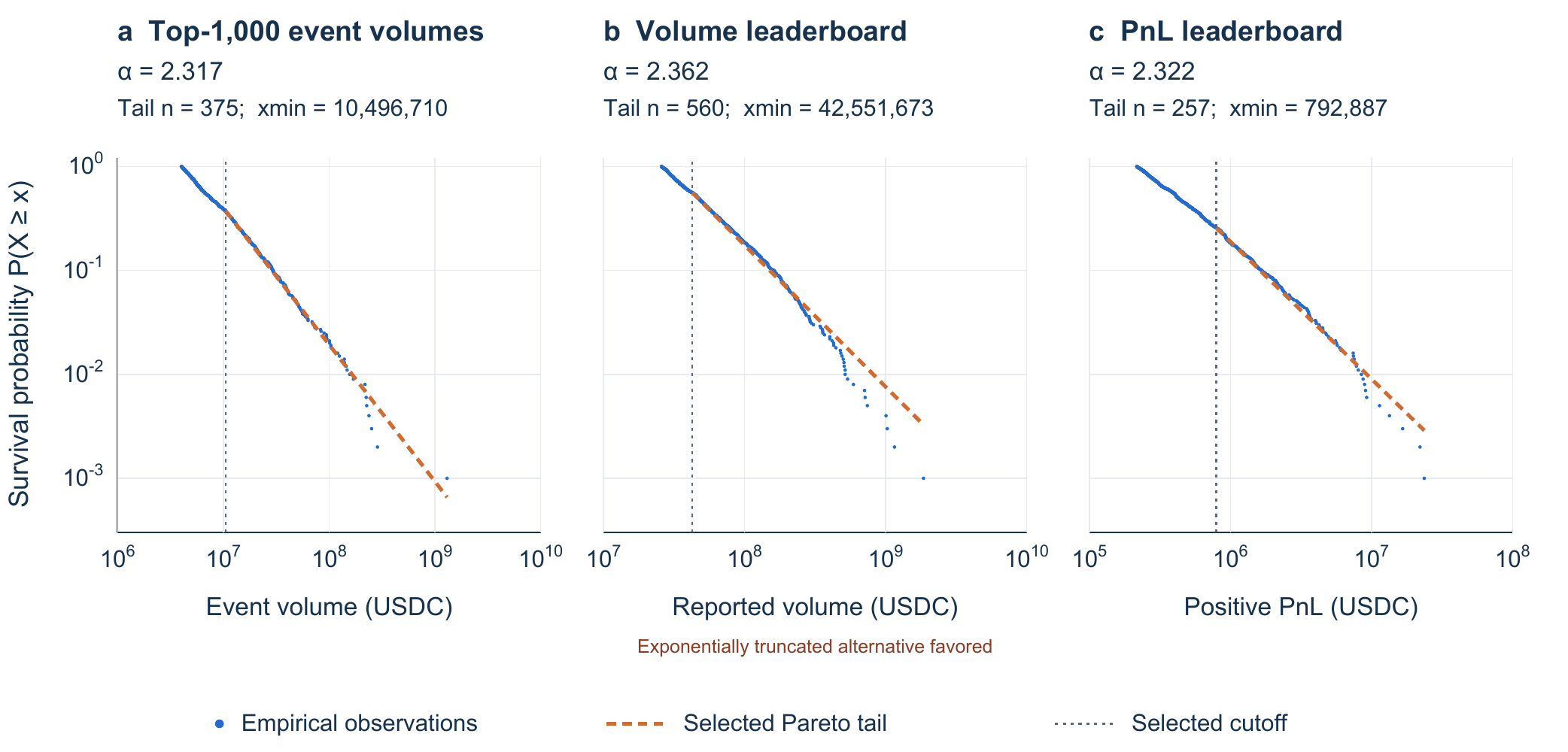}
\caption{\textbf{Trading volume and profit in three rank-selected samples.} At each amount, the blue dots show the fraction of observations at or above that value. The vertical dotted line marks the lower limit of the fitted tail ($x_{\min}$); the orange dashed line shows the fitted Pareto model. \textbf{a}, Trading volume of the 1,000 largest events. \textbf{b}, Reported volume on the all-time trading-volume leaderboard. \textbf{c}, Positive profit on the all-time profit leaderboard.}
\label{fig:tails}
\end{figure}

The reselected top-event and positive-PnL leaderboard tails are compatible with pure-Pareto fits, with exponents near 2.3 and no tested alternative favored (Table~\ref{tab:tailmain}; Fig.~\ref{fig:tails}). The volume leaderboard also meets the absolute goodness-of-fit criterion, but an exponentially truncated power law is favored on the same selected observations. Absolute compatibility and relative-model preference therefore answer different questions: non-rejection does not establish that a power law generated the data.

\begin{table}[ht]
\centering
\caption{\textbf{Principal rank-selected monetary-tail results.} Cutoffs and exponents are re-estimated by the declared CSN procedure. A value $p_{\rm KS}>0.10$ means that the fitted Pareto tail is not rejected; it is not the probability that a power law is true.}
\label{tab:tailmain}
{\small\setlength{\tabcolsep}{3.2pt}
\begin{tabular}{p{0.25\linewidth}rrrrp{0.21\linewidth}}
\toprule
Distribution & $n_t$ & $x_{\min}$ (USDC) & $\widehat\alpha$ (95\% interval) & $p_{\rm KS}$ & Relative-model result \\
\midrule
Top-1,000 event volumes & 375 & 10,496,710 & 2.317 (2.096--2.522) & 0.733 & no tested alternative favored \\
Volume leaderboard & 560 & 42,551,673 & 2.362 (2.235--2.685) & 0.154 & truncated power law favored\,$^{a}$ \\
Positive-PnL leaderboard & 257 & 792,887 & 2.322 (2.068--2.486) & 0.257 & no tested alternative favored \\
\bottomrule
\multicolumn{6}{p{0.96\linewidth}}{\footnotesize $^{a}$Exponentially truncated power law versus pure Pareto on the same selected tail: log-likelihood ratio $-4.725$, within-dataset Holm-adjusted $p=0.0063$. Intervals are bootstrap intervals for $\alpha$; they are not a common-exponent test.}
\end{tabular}}

\end{table}

These conclusions are conditional on the observed rank-selected lists. Re-estimating the cutoff within the top-event list yields a compatible tail, whereas fixing it at the list minimum does not. For the complete event-volume and event- and wallet-degree samples, the declared CSN procedure rejects the tails selected by its cutoff search. This does not test the whole distribution or rule out every possible narrower tail. Supplementary Table~S5 summarizes the eight principal fits and links to the complete results for all 60 fits, including cutoff sensitivities and model comparisons. We therefore do not describe the complete network as scale free \cite{broido}.

\subsection{Sparse participation and weight-dependent community structure}
\begin{figure}[ht]
\centering
\includegraphics[width=\textwidth]{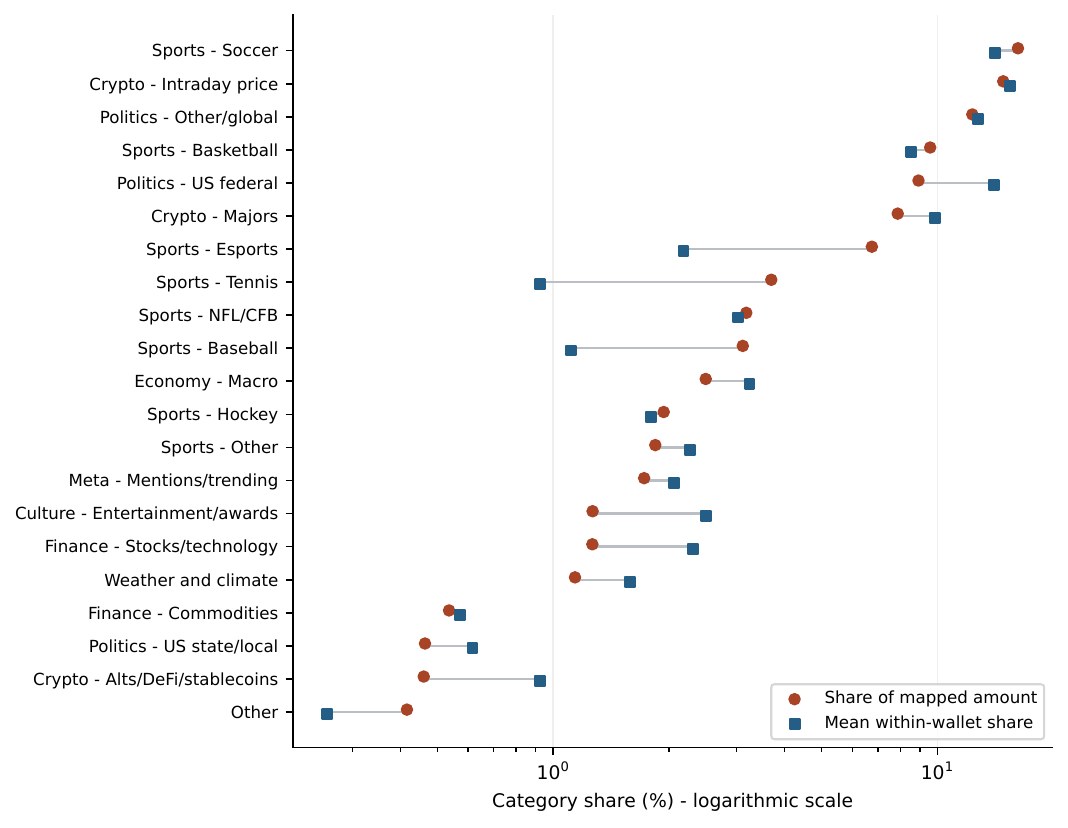}
\caption{\textbf{Subject shares when trading amounts or wallets receive equal weight.} For each of the 21 subjects, the circle shows its share of the total amount traded by all wallets. The square shows the same share calculated separately for each wallet and then averaged, giving every active wallet equal weight. A circle to the right of its square means that the subject accounts for a larger share of total trading than of the average wallet's trading. The horizontal axis is logarithmic.}
\label{fig:categoryshares}
\end{figure}

Only a small fraction of possible wallet--event pairs were linked: the bipartite density was $1.07685\times10^{-4}$. The 100 wallets with the largest traded amounts and the 100 that traded in the most events had only three addresses in common. Weighting wallets by amount rather than equally also changed aggregate subject shares (Fig.~\ref{fig:categoryshares}).

\begin{figure}[!t]
\centering
\includegraphics[width=0.88\textwidth]{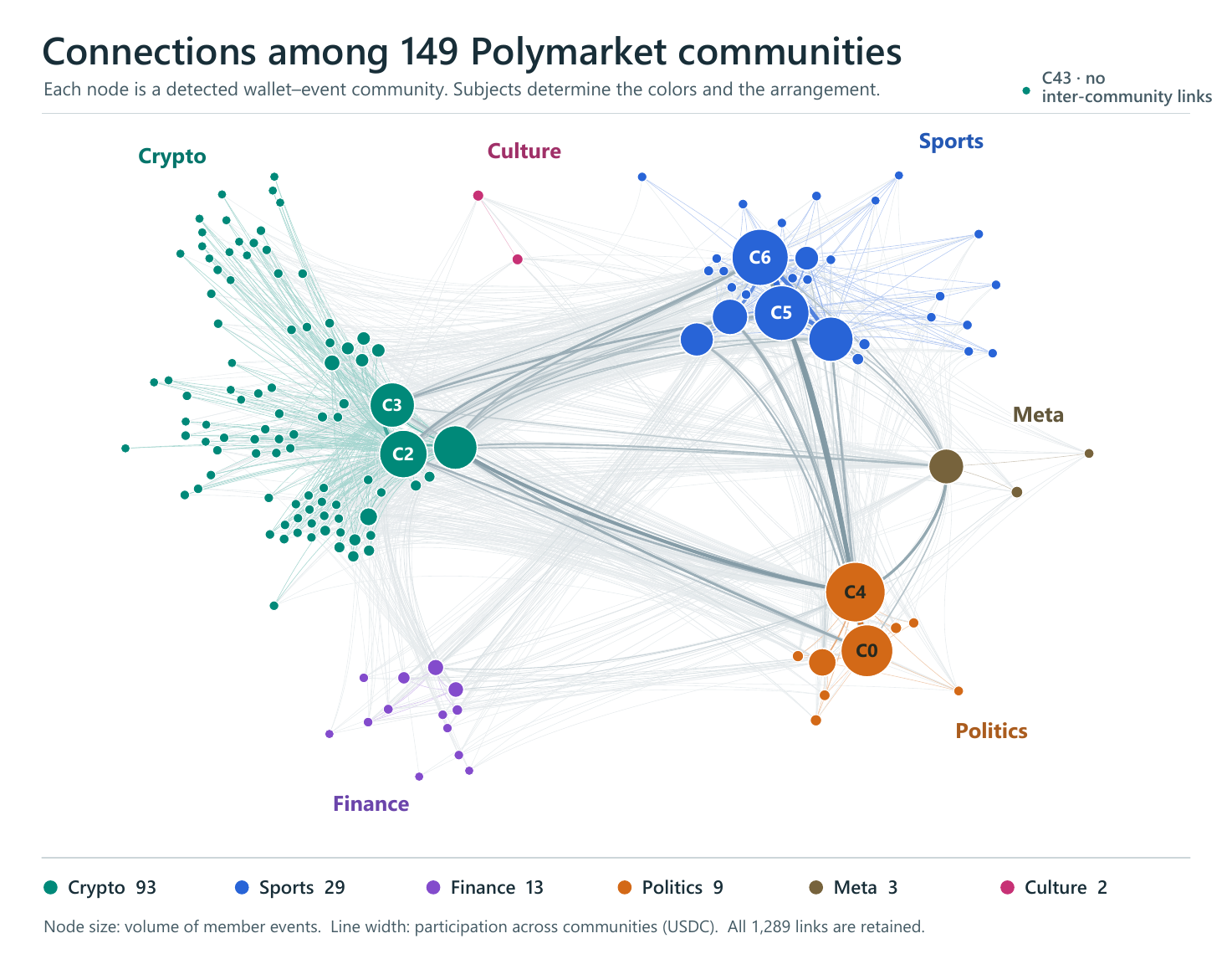}
\caption{\textbf{Trading links among the 149 Polymarket communities.}
Each node represents a community of wallets and events. Larger nodes indicate greater total trading volume in the community's events. After community detection, each node was colored by the broad family of its highest-volume detailed subject. Nodes of the same color are placed near one another for display. Link width increases with the total amount that wallets in either community traded in the other community's events. All 1,289 inter-community links are shown; C43 has no such links and is drawn separately.}
\label{fig:groupmap}
\end{figure}

We identified communities by optimizing weighted bipartite modularity, which compares trading within groups of wallets and events with the amount expected from their total activity \cite{barber,beckett}. At resolution $\gamma=1$, the amount-weighted network yielded 149 communities ($Q_b=0.6729$; 80.32\% of the weight lay within communities), compared with 3,527 when each wallet contributed equal total weight ($Q_b=0.7814$; 83.50\%). In Fig.~\ref{fig:groupmap}, each community is colored by the broad family of its leading detailed subject. We determined that subject after detection by summing the full trading volumes of the community's member events by subject and taking the largest total; no majority was required.

\subsection{Subject exposure is broad but not explained by popularity alone}
\begin{figure}[!t]
\centering
\includegraphics[width=\textwidth]{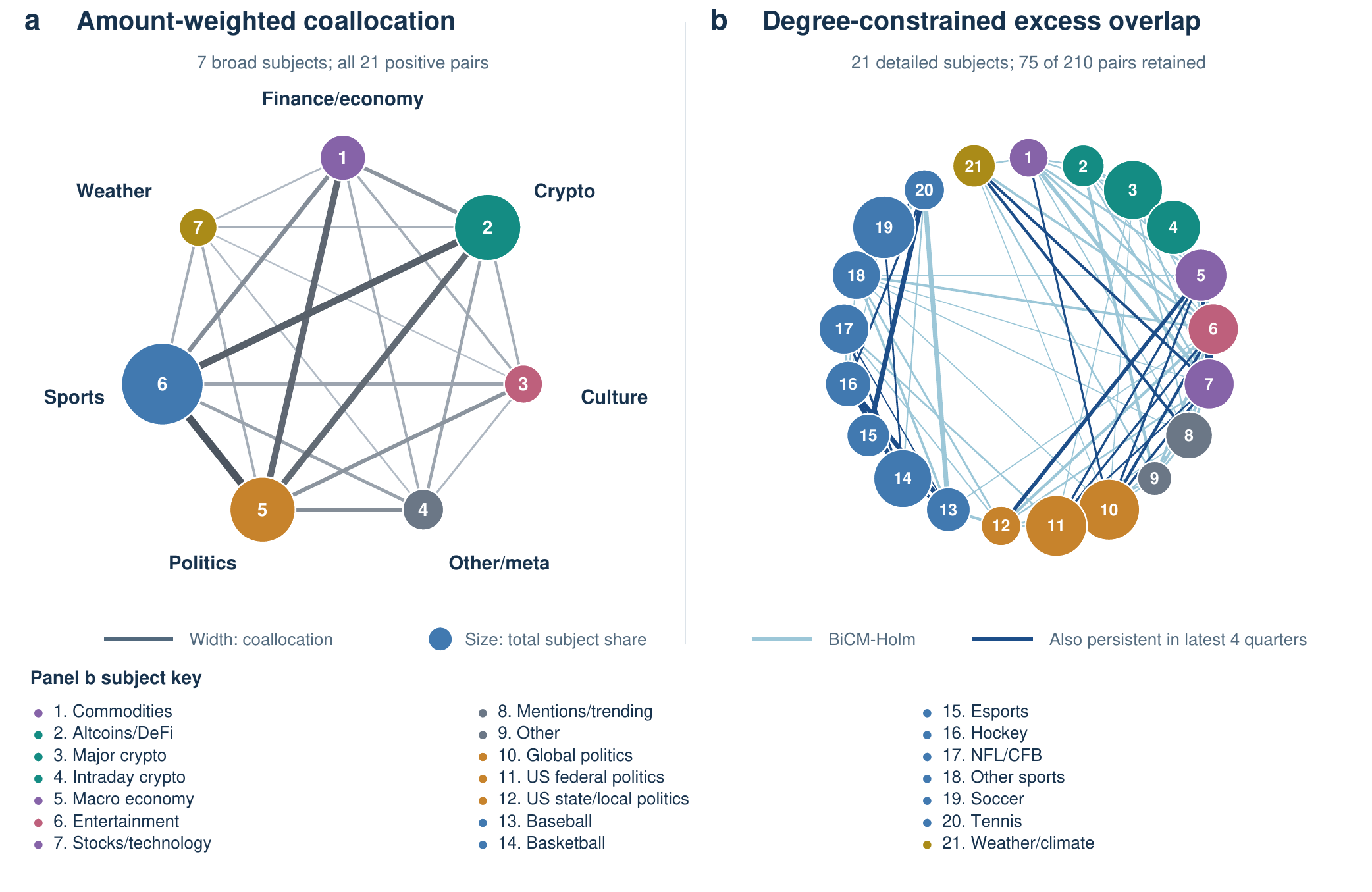}
\caption{\textbf{How wallets divide trading across subjects and which subjects share more wallets than expected.} \textbf{a}, Trading allocation across seven broad subject families. Larger nodes indicate a larger share of total trading. A link is stronger when wallets allocate substantial shares of their trading to both families; wallets that trade larger amounts carry more weight. All 21 family pairs are shown. \textbf{b}, Shared wallets across 21 detailed subjects. Larger nodes indicate more active wallets. Lines mark the 75 of 210 pairs that share more wallets than expected after accounting for each wallet's number of subjects and each subject's number of wallets (Holm-adjusted $p\leq0.05$). Thicker lines indicate a larger observed-to-expected ratio. Dark lines mark the 22 pairs for which this ratio exceeded one in each of the last four complete quarters.}
\label{fig:subjectnetworks}
\end{figure}

The subject networks in Fig.~\ref{fig:subjectnetworks} answer two different questions. The amount-weighted coallocation view collapses the 21 detailed classes into seven broad families and weights each wallet by its total participation amount. All 21 broad-family pairs have positive coallocation, with the strongest links concentrated around Sports, Politics and Crypto. This is an allocation map, not a significance network.

The second view uses the binary wallet--subject matrix and a fitted bipartite configuration model (BiCM). Every one of the 210 detailed-subject pairs shares wallets, but only 75 pairs have more shared wallets than expected after controlling, in expectation, for each wallet's subject breadth and each subject's wallet count, followed by Holm correction. Twenty-two retained pairs also have observed-to-expected overlap above one in each of the latest four complete quarters. This persistence check records direction, not four additional significance tests.

An exploratory Sports/non-Sports analysis finds more within-block than cross-block enrichment. In the amount projection, positive lift relative to a strength-product reference occurs for 24 of 28 Sports--Sports pairs, 63 of 78 other within-block pairs and 2 of 104 cross-block pairs. The corresponding BiCM-Holm counts are 16, 49 and 10. This split was examined after the pattern was noticed; its exact enumeration results are post hoc diagnostics (Supplementary Section~S5.3 and Fig.~S1), not confirmatory evidence of a prespecified division.

\subsection{High-purchase candidate cohorts have more similar subject allocations}
The presidential-election case study separates formation from follow-up. September 2024 taker-side YES purchases in the focal presidential-election winner event define Harris-only and Trump-only groups. Purchase-size quantiles are calculated across all 21,301 buyers of either named outcome, including buyers of both, before selecting the exclusive cohorts. Bottom 5\% denotes at most 1.09414 USDC and Top 5\% at least 2,787.73554 USDC, retaining boundary ties. These labels describe observed focal-market purchases; they do not measure personal wealth or political affiliation.

The eligible cohort sizes are 504 Harris-Bottom (HB), 541 Trump-Bottom (TB), 334 Harris-Top (HT) and 251 Trump-Top (TT). October non-focal activity is observed for 360, 308, 168 and 135 wallets, respectively. October was chosen as the final complete calendar month after September formation and before the 5 November election; the focal event is excluded from follow-up. This separates formation from measurement without using a partial November window. The comparison remains conditional on October activity, with unequal active fractions, and does not establish that October represents other months.

\begin{figure}[!p]
\centering
\includegraphics[width=\textwidth]{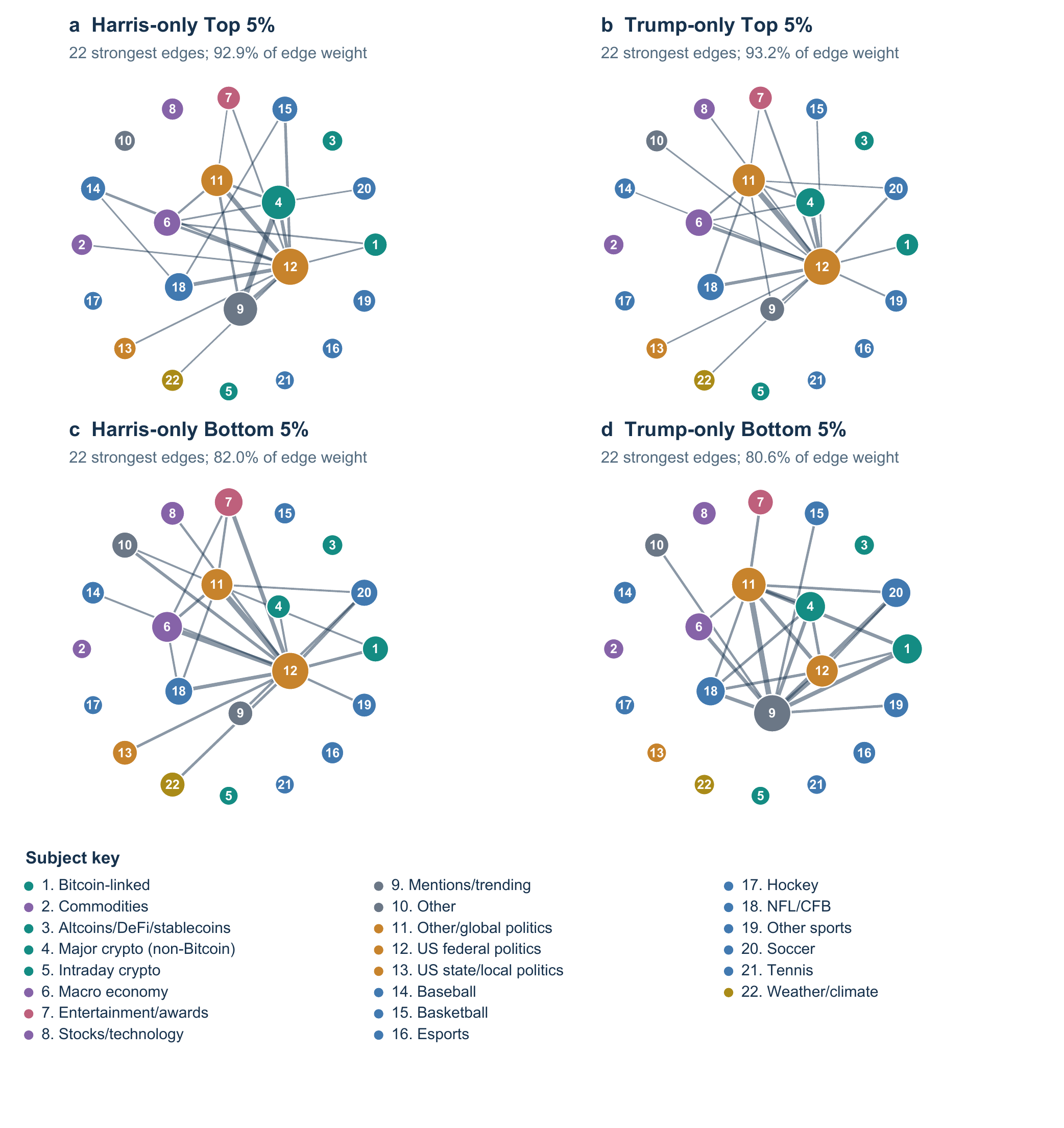}
\caption{\textbf{October trading across subjects by wallets that backed Harris or Trump in September.} Columns distinguish wallets that bought YES on Harris but not Trump (left) from those that bought YES on Trump but not Harris (right). Rows show the top 5\% (\textbf{a,b}) and bottom 5\% (\textbf{c,d}) by September election purchase amount. Nodes are subjects; a link indicates that wallets traded in both subjects outside the focal election event. Node size reflects the combined weight of a subject's links, and line width reflects how strongly wallets allocated trading to both subjects. Only the 22 strongest links are shown; the percentage above each panel is their share of total link weight. Subject positions are fixed across panels, but sizes and widths are scaled within each panel and do not compare dollar amounts across groups.}
\label{fig:electionnetworks}
\end{figure}

Figure~\ref{fig:electionnetworks} displays the 22 strongest amount-coallocation edges per cohort at a common edge budget. The displayed edges contain 82.0\%, 80.6\%, 92.9\% and 93.2\% of complete off-diagonal edge mass for HB, TB, HT and TT. Ninety per cent of edge mass requires 37 and 35 edges in the Bottom networks but 17 in each Top network. The high-purchase networks are therefore more concentrated around a smaller set of subject links.

The more concentrated Top networks also have more similar allocations across candidate cohorts. After normalizing each complete amount-edge matrix to unit off-diagonal mass, the Top Harris--Trump pair has lower Jensen--Shannon divergence (JSD) than the Bottom pair: 0.243 versus 0.457 bits. This comparison uses all 231 subject pairs, not only the displayed edges (Supplementary Table~S8).

The Top Harris--Trump pair is closer than the Bottom pair in all six normalized profile and edge representations. Amount-profile JSD is 0.278 bits for the Bottom pair and 0.073 bits for the Top pair; corresponding Bottom/Top values are 0.061/0.013 for prevalence profiles and 0.186/0.074 for binary edges. Across all six cohort pairs, HT--TT is the closest pair in five representations; the exception is the amount-edge view, where HB--TT is closer. Thus, the high-purchase similarity is not confined to dollar-weighted links, although individual monetary comparisons differ (Supplementary Tables~S8 and S10).

Within-cohort wallet bootstrapping gives positive pointwise 95\% intervals for Bottom-minus-Top differences in amount profiles, prevalence, binary edges and, narrowly, equal-wallet edges; the other two intervals include zero (Supplementary Table~S8). A separate fixed-vector randomization reassigns candidate labels within each purchase tier while preserving the observed wallet vectors and active cohort sizes. After Holm correction across six directional contrasts, differences are supported for equal-wallet profiles, prevalence profiles, equal-wallet edges and binary edges (adjusted $p=0.0006$ for each), but not for amount profiles ($p=0.3274$) or amount edges ($p=0.3357$; Supplementary Table~S9). Bootstrap uncertainty concerns sampling within the observed cohorts; label randomization instead tests whether the candidate-label ordering is unusual conditional on the tier-specific wallet vectors.

Top-five hub Jaccard similarity, the intersection divided by the union of the hub sets, is 0.667 for the Top pair versus 0.250 for the Bottom pair in amount and equal-wallet views; binary hub similarity is 1.000 versus 0.429. The 22 strongest amount edges have Jaccard similarity 0.571 in the Top pair and 0.333 in the Bottom pair (Supplementary Section~S6.4).

These comparisons concern cohort-aggregate profiles and networks, not similarity among individual wallets within a cohort. Similar aggregate allocation also does not imply matching communities of subject nodes: cross-cohort adjusted Rand indices of the best subject partitions are low under amount, equal-wallet and binary representations. A strength-ranked subject-core permutation diagnostic is significant in all four networks, but its largest enrichment occurs in TB rather than in a Top cohort. These comparisons identify shared profiles, hubs and dominant links, not a common modular partition or a wallet rich club.

\FloatBarrier

\section{Discussion}
Trading on Polymarket is not simply a matter of expressing an opinion. Users put money behind their predictions and accept financial risk. In this way, they reveal beliefs that are different in nature from opinions expressed without any cost. This mechanism can be considered a form of ``costly belief,'' since users do not only express their views but also put their money at risk. However, accepting financial risk does not necessarily mean that these users have better information or can make more accurate predictions. In this study, using approximately 1.14 billion transaction records, more than 3.1 million wallet addresses, and 764,000 prediction events, we analyze the participation structure of Polymarket from three perspectives: financial concentration, network organization, and participation patterns related to political choices.

Our results show that participation in Polymarket is highly unequal: the top 1\% of events account for 64\% of the total transaction volume, while the top 1\% of wallets account for 79.2\% of financial participation. This concentration is also reflected in the Gini coefficients, which are 0.933 for events and 0.962 for wallets. The analysis of the distributions shows that heavy-tailed behavior and compatibility with a Pareto distribution are limited to specific parts of the data. 
Across the full wallet--event network, only three of the 100 wallets with the largest total trading amounts were also among the 100 that traded in the most distinct events.

The results provide an interesting and somewhat different picture. Among wallets in the top 5\% of YES purchases, the patterns of participation allocation across subjects outside the focal election event are considerably more similar between the Trump-only and Harris-only groups. In contrast, these patterns are more diverse and divergent among wallets in the bottom 5\% of purchases. This result is observed across all six representations considered in the study, including amount and participation-prevalence profiles, as well as edge-based representations. However, the permutation test with multiple-testing correction provides statistical support for four of these contrasts (with the smallest adjusted \(p\)-value equal to 0.0006). In other words, even though the two groups made different election-related choices, their allocation patterns across other subjects are more similar among wallets with higher financial participation. In contrast, wallets with lower financial participation show more diverse patterns.

The Harris- and Trump-linked groups allocated their trading across subjects more similarly at high election stakes than at low stakes. This pattern is consistent with the idea of ``convergence of costly beliefs'': political preferences may leave more room for differences in allocation when less money is at stake. The comparison, however, is between groups. Their similarity at high stakes does not imply that individual traders agreed or were better informed. Nor can we attribute the larger difference at low stakes to political bias: wallet addresses need not represent distinct individuals, and an election purchase does not establish the buyer's political affiliation or motive.

\section{Methods}

\subsection{Concentration measures}
For positive amounts $x_{(1)}\leq\cdots\leq x_{(N)}$, the empirical Lorenz curve joins $(0,0)$ to
\[
\left(k/N,\;\sum_{i=1}^{k}x_{(i)}/\sum_{i=1}^{N}x_{(i)}\right),\quad k=1,\ldots,N,
\]
and the Gini coefficient is $2\sum_i i x_{(i)}/(N\sum_i x_{(i)})-(N+1)/N$, twice the area between the curve and the diagonal of perfect equality \cite{lorenz,gastwirth}. Upper shares use the largest $\lceil qN\rceil$ observations and make no Pareto assumption.

\subsection{Power-law analysis}
For continuous values $x\geq x_{\min}$, the candidate density is
\begin{equation}
f(x)=(\alpha-1)x_{\min}^{\alpha-1}x^{-\alpha},\qquad
\widehat\alpha=1+\frac{n_t}{\sum_{x_i\geq x_{\min}}\log(x_i/x_{\min})}.
\label{eq:pareto}
\end{equation}
The analysis follows the CSN sequence: maximum-likelihood estimation, cutoff selection by minimum two-sided Kolmogorov--Smirnov distance, and semiparametric goodness-of-fit simulation with refitting in every synthetic sample \cite{clauset,newmanpower}. Tails contain at least 100 observations. The three main $n=1{,}000$ datasets use exhaustive search over 901 admissible cutoffs; larger continuous arrays use a union of 500 linearly and 500 geometrically spaced tail-size targets. This grid is a declared computational approximation, not an exhaustive cutoff search for the largest arrays. Integer degrees use the exact discrete Hurwitz-zeta likelihood.

Each specified distribution uses 2,500 valid goodness-of-fit simulations, except wallet degree (10,000). Synthetic samples combine a resampled empirical body with a generated Pareto tail, allowing the tail count to vary binomially while retaining the original sample size. Both cutoff and exponent are re-estimated, except in the explicitly fixed-cutoff top-1,000 sensitivity, which refits only the exponent. Reported probabilities use $(K_{\geq}+1)/(B+1)$, where $K_{\geq}$ is the number of simulated KS distances at least as large as observed \cite{phipson}. A pure Pareto candidate is rejected when $p_{\rm KS}\leq0.10$. For 58 continuous candidates, the fitted Pareto is compared on the same selected tail with exponential, lognormal, stretched-exponential and exponentially truncated power-law alternatives. Non-nested comparisons use signed likelihood ratios \cite{vuong}; within each candidate, Holm correction covers the four alternatives \cite{holm}. The two discrete degree candidates receive absolute-fit tests but no alternative-family comparison. The three displayed exponent intervals instead use 1,000 nonparametric bootstrap resamples of each observed dataset, reselecting the cutoff and refitting the exponent. Pointwise survival plots do not enter estimation.

\subsection{Bipartite modularity and community detection}
For either weighting defined in Section~\ref{sec:participation-data}, let $w_{ue}$ denote the chosen edge weight, $s_u=\sum_e w_{ue}$ the wallet strength and $s_e=\sum_u w_{ue}$ the event strength. Weighted bipartite modularity is
\begin{equation}
Q_b=\frac{1}{m}\sum_{u,e}\left(w_{ue}-\gamma\frac{s_us_e}{m}\right)\mathbf{1}\{g_u=g_e\},\qquad m=\sum_{u,e}w_{ue}.
\label{eq:barber}
\end{equation}
The null term preserves strengths in expectation \cite{barber,beckett}. The memory-bounded search alternates original-node moves with directed Leiden optimization of the exact aggregated quotient and is repeated across declared seeds and resolutions \cite{leichtnewman,traag}. Numerical checks verify that aggregation preserves the objective. Subject and quarter labels are attached only after optimization.

\subsection{Subject coallocation and degree-constrained overlap}
Let $x_{uc}$ be wallet $u$'s amount in subject $c$, $v_u=\sum_c x_{uc}$ and $p_{uc}=x_{uc}/v_u$. The amount- and equal-wallet coallocation matrices are
\begin{equation}
W^{A}_{cd}=\frac{\sum_u v_up_{uc}p_{ud}}{\sum_u v_u},\qquad
W^{E}_{cd}=\frac{1}{n}\sum_u p_{uc}p_{ud},\quad c\ne d.
\label{eq:subjectprojection}
\end{equation}
The seven-family panel sums the 21-class matrices according to a fixed mapping; no edge is filtered. Binary shared-wallet overlap $V_{cd}=\sum_u A_{uc}A_{ud}$ is tested against a BiCM with independent probabilities $P(A_{uc}=1)=x_uy_c/(1+x_uy_c)$, with multipliers fitted to reproduce wallet and subject degrees in expectation \cite{saracco}. For each of 210 pairs, 19,999 marginal shared-count simulations yield a plus-one probability of simulated overlap at least as large as observed. Holm correction covers the 210-pair family; edges are retained at adjusted $p\leq0.05$. Pointwise Clopper--Pearson limits quantify Monte Carlo error but are not simultaneous confidence intervals \cite{clopperpearson}. A projected edge means shared participation, not communication or a social tie \cite{nealgroup,peel}.

\subsection{Election cohort formation and October measurements}
September 2024 taker-side YES purchases in the focal presidential-election winner event define Harris-only wallets that bought the Harris outcome but not Trump, and Trump-only wallets with the reverse pattern. The quantile reference pool contains all 21,301 buyers of either outcome: 7,474 Harris-only, 9,361 Trump-only and 4,466 both-outcome wallets. Cutoffs use total named-outcome YES purchase amount and retain boundary ties; only then are the exclusive cohorts selected. October measurements exclude the focal event and use the original taxonomy with a title-audited 21-event Bitcoin subdivision, producing 22 mutually exclusive subjects.

For each October-active wallet $u$, let $x_{ui}$ be amount in subject $i$, $v_u=\sum_i x_{ui}$ and $p_{ui}=x_{ui}/v_u$. Three complete off-diagonal subject matrices are
\begin{align}
W^{A}_{ij}&=\frac{\sum_u v_up_{ui}p_{uj}}{\sum_u v_u}, &
W^{E}_{ij}&=\frac{1}{n}\sum_u p_{ui}p_{uj}, &
W^{B}_{ij}&=\frac{1}{n}\sum_u \mathbf{1}(x_{ui}>0)\mathbf{1}(x_{uj}>0).
\label{eq:electionnetworks}
\end{align}
The corresponding node profiles are pooled amount share, mean within-wallet share and subject prevalence. Each edge vector is normalized to unit sum before comparison.

For normalized vectors $P$ and $Q$, JSD in bits is $\tfrac12D_{\rm KL}(P\|M)+\tfrac12D_{\rm KL}(Q\|M)$, where $M=(P+Q)/2$ and logarithms have base two \cite{linjsd}. Six-view similarity is $1$ minus the unweighted mean of three profile and three edge JSDs. The display uses the strongest 22 edges per cohort under fixed node positions; every reported divergence uses the complete matrix. Edge concentration is summarized by $D_2^{\rm edge}=1/\sum_e q_e^2$, the number of equally weighted edges with the same concentration \cite{hilldiversity}.

\subsection{Statistical analysis}
Election Bottom-minus-Top JSD uncertainty is estimated from 1,999 within-cohort wallet bootstrap replicates \cite{efron}. Each draw samples, with replacement, the original number of eligible September-cohort wallets, including zero-October vectors. Representations are reconstructed from the sampled positive-October rows, so follow-up activity counts can vary. The six 2.5th/97.5th percentile intervals are pointwise and unadjusted across correlated views. In a retrospective sensitivity analysis, the directional contrast $\Delta={\rm JSD}_{\rm Bottom}-{\rm JSD}_{\rm Top}>0$ is tested by 9,999 fixed-vector randomizations that independently reassign candidate labels within each purchase tier while preserving active cohort sizes; the same assignment is used for all six representations in a replicate. Plus-one upper-tail probabilities are Holm-adjusted across the six views. This null conditions on the observed tier-specific wallet-vector multisets, including spending, breadth and allocation concentration, while assuming exchangeability of candidate labels and independence of wallet rows. Weighted Louvain optimization at resolution one is repeated over 200 seeds only as a sensitivity to modular subdivision \cite{blondel}. The Sports/non-Sports split is post hoc. Supplementary Table~S11 distinguishes resampling, null simulation and optimization repeats; Sections~S3--S7 specify their statistics, populations and settings.

\section*{Data availability}
Trade records are available from TimeSeventeen's public Polymarket-v1 and Polymarket-v2 datasets on Hugging Face \cite{source,sourcev2}; Qin and Yang describe v1 \cite{qin}. We used the cleaned Parquet trade layers (\texttt{daily\_aligned} and \texttt{daily\_aligned\_multi}) at the revisions linked in Supplementary Table~S1. We did not obtain fills directly from Polygon nodes or from Polymarket's order-book trade history.

For this study we collected event and market metadata from the Gamma API and leaderboard observations from the Data API. The leaderboard request (\texttt{GET /v1/leaderboard}) used \texttt{category=OVERALL} and \texttt{timePeriod=ALL}, ordered separately by \texttt{VOL} and \texttt{PNL}. The study archive contains the metadata fields used for event mapping and classification, the two dated ranking series, numerical inputs for the figures, and file checksums. It does not redistribute the third-party fill files or the full wallet--event edge list.
\ifdefined\StudyReleaseURL The versioned code and data are available in the \href{\StudyReleaseURL}{study repository}.\fi

\section*{Code availability}
The study archive contains the event classifier, analysis code and scripts for reproducing the figures from their numerical inputs. Reconstructing the full wallet--event network requires the original fill files and the dated condition-to-event mapping; the aggregate figure inputs do not contain individual edges or node assignments (Supplementary Section~S7).




\section*{Additional information}
Supplementary Information accompanies this paper.



\begin{thebibliography}{99}
\small
\interlinepenalty=10000
\setlength{\itemsep}{3pt}
\setlength{\parsep}{0pt}
\setlength{\parskip}{0pt}
\raggedright

\bibitem{wolfers}
Wolfers J, Zitzewitz E. Prediction markets. \emph{Journal of Economic Perspectives}. 2004;18(2):107--126.
\href{https://doi.org/10.1257/0895330041371321}{doi:10.1257/0895330041371321}.
\bibitem{rothschildsethi}
Rothschild DM, Sethi R. Trading strategies and market microstructure: evidence from a prediction market. \emph{Journal of Prediction Markets}. 2016;10(1):1--29.
\href{https://doi.org/10.5750/jpm.v10i1.1179}{doi:10.5750/jpm.v10i1.1179}.
\bibitem{tsang}
Tsang KP, Yang Z. The anatomy of a blockchain prediction market: Polymarket in the 2024 U.S. presidential election. 2026. Preprint, arXiv:2603.03136v3.
\href{https://doi.org/10.48550/arXiv.2603.03136}{doi:10.48550/arXiv.2603.03136}.
\bibitem{sokdpm}
Rahman N, Al-Chami J, Clark J. SoK: market microstructure for decentralized prediction markets (DePMs). Presented at the Workshop on Trusted Smart Contracts, Financial Cryptography and Data Security; 2026. arXiv:2510.15612.
\href{https://doi.org/10.48550/arXiv.2510.15612}{doi:10.48550/arXiv.2510.15612}.
\bibitem{saguillo}
Saguillo O, Ghafouri V, Kiffer L, Suarez-Tangil G. Unravelling the probabilistic forest: arbitrage in prediction markets. In: \emph{7th Conference on Advances in Financial Technologies (AFT 2025)}. LIPIcs. 2025;354:27:1--27:24.
\href{https://doi.org/10.4230/LIPIcs.AFT.2025.27}{doi:10.4230/LIPIcs.AFT.2025.27}.
\bibitem{tsangshocks}
Tsang KP, Yang Z. Political shocks and price discovery in prediction markets: evidence from the 2024 U.S. presidential election. 2026. Preprint, arXiv:2603.03152v4.
\href{https://doi.org/10.48550/arXiv.2603.03152}{doi:10.48550/arXiv.2603.03152}.
\bibitem{qin}
Qin B, Yang R. Polymarket-v1 database. 2026. Preprint, arXiv:2606.04217v2.
\href{https://doi.org/10.48550/arXiv.2606.04217}{doi:10.48550/arXiv.2606.04217}.
\bibitem{akey}
Akey P, Gr\'egoire V, Harvie N, Martineau C. Who wins and who loses in prediction markets? Evidence from Polymarket. CEPR Discussion Paper 21615; 2026. Also available as SSRN 6443103.
\href{https://doi.org/10.2139/ssrn.6443103}{doi:10.2139/ssrn.6443103}.
\bibitem{chenweb3}
Chen H, Duan X, El Saddik A, Cai W. Political leanings in Web3 betting: decoding the interplay of political and profitable motives. In: \emph{Proceedings of the 17th ACM Web Science Conference}. 2025. pp. 96--105.
\href{https://doi.org/10.1145/3717867.3717884}{doi:10.1145/3717867.3717884}.
\bibitem{gomezcram}
G\'omez-Cram R, Guo Y, Jensen TI, Kung H. Prediction market accuracy: crowd wisdom or informed minority? 2026. Working paper.
\href{https://doi.org/10.2139/ssrn.6617059}{doi:10.2139/ssrn.6617059}.
\bibitem{luongheesen}
Luong KL, Heesen G. The wisdom of the few: skilled traders and prediction market accuracy. 2026. Working paper, SSRN 6758662.
\href{https://doi.org/10.2139/ssrn.6758662}{doi:10.2139/ssrn.6758662}.
\bibitem{farmer2002ecology}
Farmer JD. Market force, ecology and evolution.
\emph{Industrial and Corporate Change}. 2002;11(5):895--953.
\href{https://doi.org/10.1093/icc/11.5.895}{doi:10.1093/icc/11.5.895}.

\bibitem{scholl2021ecology}
Scholl MP, Calinescu A, Farmer JD. How market ecology explains market malfunction.
\emph{Proceedings of the National Academy of Sciences}. 2021;118(26):e2015574118.
\href{https://doi.org/10.1073/pnas.2015574118}{doi:10.1073/pnas.2015574118}.
\bibitem{lorenz}
Lorenz MO. Methods of measuring the concentration of wealth. \emph{Publications of the American Statistical Association}. 1905;9(70):209--219.
\href{https://doi.org/10.1080/15225437.1905.10503443}{doi:10.1080/15225437.1905.10503443}.
\bibitem{gastwirth}
Gastwirth JL. The estimation of the Lorenz curve and Gini index. \emph{The Review of Economics and Statistics}. 1972;54(3):306--316.
\href{https://doi.org/10.2307/1937992}{doi:10.2307/1937992}.
\bibitem{newmanpower}
Newman MEJ. Power laws, Pareto distributions and Zipf's law. \emph{Contemporary Physics}. 2005;46(5):323--351.
\href{https://doi.org/10.1080/00107510500052444}{doi:10.1080/00107510500052444}.
\bibitem{broido}
Broido AD, Clauset A. Scale-free networks are rare. \emph{Nature Communications}. 2019;10:1017.
\href{https://doi.org/10.1038/s41467-019-08746-5}{doi:10.1038/s41467-019-08746-5}.
\bibitem{source}
TimeSeventeen. Polymarket-v1 [dataset]. Hugging Face.
\url{https://huggingface.co/datasets/TimeSeventeen/Polymarket-v1/tree/5aa1b9d52316a8b2e789e81c8ae42c7ed532e8aa}.
Version-specific public dataset; accessed 1 September 2026.
\bibitem{sourcev2}
TimeSeventeen. Polymarket-v2 [dataset]. Hugging Face.
\url{https://huggingface.co/datasets/TimeSeventeen/Polymarket-v2/tree/2a8dba6472f4c9ca558869e9cc1a65b9e1614264}.
Version-specific public dataset; accessed 1 September 2026.
\bibitem{gammaapi}
Polymarket. List events (keyset pagination) [Gamma API documentation].
\url{https://docs.polymarket.com/api-reference/events/list-events-keyset-pagination}.
Accessed 13 September 2026; metadata collection dated 1 September 2026.
\bibitem{leaderboardapi}
Polymarket. Get trader leaderboard rankings [Data API v1 documentation].
\url{https://docs.polymarket.com/api-reference/core/get-trader-leaderboard-rankings}.
Accessed 13 September 2026; leaderboard snapshot dated 4 September 2026.
\bibitem{clauset}
Clauset A, Shalizi CR, Newman MEJ. Power-law distributions in empirical data. \emph{SIAM Review}. 2009;51(4):661--703.
\href{https://doi.org/10.1137/070710111}{doi:10.1137/070710111}.
\bibitem{phipson}
Phipson B, Smyth GK. Permutation $p$-values should never be zero: calculating exact $p$-values when permutations are randomly drawn. \emph{Statistical Applications in Genetics and Molecular Biology}. 2010;9(1):Article 39.
\href{https://doi.org/10.2202/1544-6115.1585}{doi:10.2202/1544-6115.1585}.
\bibitem{vuong}
Vuong QH. Likelihood ratio tests for model selection and non-nested hypotheses. \emph{Econometrica}. 1989;57(2):307--333.
\href{https://doi.org/10.2307/1912557}{doi:10.2307/1912557}.
\bibitem{holm}
Holm S. A simple sequentially rejective multiple test procedure. \emph{Scandinavian Journal of Statistics}. 1979;6(2):65--70.
\bibitem{barber}
Barber MJ. Modularity and community detection in bipartite networks. \emph{Physical Review E}. 2007;76:066102.
\href{https://doi.org/10.1103/PhysRevE.76.066102}{doi:10.1103/PhysRevE.76.066102}.
\bibitem{beckett}
Beckett SJ. Improved community detection in weighted bipartite networks. \emph{Royal Society Open Science}. 2016;3(1):140536.
\href{https://doi.org/10.1098/rsos.140536}{doi:10.1098/rsos.140536}.
\bibitem{leichtnewman}
Leicht EA, Newman MEJ. Community structure in directed networks. \emph{Physical Review Letters}. 2008;100:118703.
\href{https://doi.org/10.1103/PhysRevLett.100.118703}{doi:10.1103/PhysRevLett.100.118703}.
\bibitem{traag}
Traag VA, Waltman L, van Eck NJ. From Louvain to Leiden: guaranteeing well-connected communities. \emph{Scientific Reports}. 2019;9:5233.
\href{https://doi.org/10.1038/s41598-019-41695-z}{doi:10.1038/s41598-019-41695-z}.
\bibitem{saracco}
Saracco F, Di Clemente R, Gabrielli A, Squartini T. Randomizing bipartite networks: the case of the World Trade Web. \emph{Scientific Reports}. 2015;5:10595.
\href{https://doi.org/10.1038/srep10595}{doi:10.1038/srep10595}.
\bibitem{clopperpearson}
Clopper CJ, Pearson ES. The use of confidence or fiducial limits illustrated in the case of the binomial. \emph{Biometrika}. 1934;26(4):404--413.
\href{https://doi.org/10.1093/biomet/26.4.404}{doi:10.1093/biomet/26.4.404}.
\bibitem{nealgroup}
Neal ZP. When can networks be inferred from observed groups? \emph{Network Science}. 2024;12:189--200.
\href{https://doi.org/10.1017/nws.2024.6}{doi:10.1017/nws.2024.6}.
\bibitem{peel}
Peel L, Peixoto TP, De Domenico M. Statistical inference links data and theory in network science. \emph{Nature Communications}. 2022;13:6794.
\href{https://doi.org/10.1038/s41467-022-34267-9}{doi:10.1038/s41467-022-34267-9}.
\bibitem{linjsd}
Lin J. Divergence measures based on the Shannon entropy. \emph{IEEE Transactions on Information Theory}. 1991;37:145--151.
\href{https://doi.org/10.1109/18.61115}{doi:10.1109/18.61115}.
\bibitem{hilldiversity}
Hill MO. Diversity and evenness: a unifying notation and its consequences. \emph{Ecology}. 1973;54(2):427--432.
\href{https://doi.org/10.2307/1934352}{doi:10.2307/1934352}.
\bibitem{efron}
Efron B. Bootstrap methods: another look at the jackknife. \emph{The Annals of Statistics}. 1979;7(1):1--26.
\href{https://doi.org/10.1214/aos/1176344552}{doi:10.1214/aos/1176344552}.
\bibitem{blondel}
Blondel VD, Guillaume JL, Lambiotte R, Lefebvre E. Fast unfolding of communities in large networks. \emph{Journal of Statistical Mechanics: Theory and Experiment}. 2008;2008(10):P10008.
\href{https://doi.org/10.1088/1742-5468/2008/10/P10008}{doi:10.1088/1742-5468/2008/10/P10008}.
\end{thebibliography}
\end{document}